\documentclass[11pt]{article}
\usepackage{amsmath}
\usepackage{amsfonts}
\usepackage{amssymb}
\usepackage{graphicx}
\usepackage{bm}
\usepackage{color}
\usepackage{amscd}

\def\bea{\begin{eqnarray}}
\def\eea{\end{eqnarray}}

\begin{document}
\begin{center}
\LARGE {\bf Conserved Charges of Minimal Massive Gravity Coupled to Scalar Field }
\end{center}

\begin{center}
{M. R. Setare \footnote{E-mail: rezakord@ipm.ir}\hspace{1mm} ,
H. Adami \footnote{E-mail: hamed.adami@yahoo.com}\hspace{1.5mm} \\
{\small {\em  Department of Science, University of Kurdistan, Sanandaj, Iran.}}}\\

\end{center}

\begin{center}
{\bf{Abstract}}\\
Recently, the theory of Topologically massive gravity non-minimally coupled to a scalar field has been proposed which comes from Lorentz-Chern-Simons theory \cite{1}. That theory is a torsion free one. We extend that theory by adding an extra term which makes torsion to be non-zero. The extended theory can be regarded as an extension of Minimal massive gravity such that it is non-minimally coupled to a scalar field. We obtain equations of motion of extended theory such that they are expressed in terms of usual torsion free spin-connection. We show that BTZ spacetime is a solution of this theory when scalar field is constant. We define quasi-local conserved charge by the concept of generalized off-shell ADT current which both are conserved for any asymptotically Killing vector field as well as a Killing vector field which is admitted by spacetime everywhere. Also we find general formula for entropy of stationary black hole solution in the context of considered theory. We apply the obtained formulas on BTZ black hole solution, to calculate energy, angular momentum and entropy of this soultion. We obtain the central extension term, the central charges and the eigenvalues of the Virasoro algebra generators for the BTZ black hole solution. We calculate energy and angular momentum of BTZ black hole using the eigenvalues of the Virasoro algebra generators.  Also we calculate the entropy of BTZ black hole by using the Cardy formula. We find that obtained results using two different ways are exactly matched as we expected.
\end{center}

\section{Introduction}
 We know that the pure Einstein-–Hilbert gravity in three dimensions exhibits no propagating physical degrees of freedom \cite{2',3}. But adding the gravitational Chern-Simons term produces a
propagating massive graviton \cite{4}. The resulting theory
is called topologically massive gravity (TMG). Including a negative cosmological constant, yields cosmological topologically massive
 gravity (CTMG). In this case the theory exhibits both gravitons and black holes.  Unfortunately there is a problem in this model, with the usual
sign for the gravitational constant, the massive excitations of CTMG carry negative energy. In the absence of a cosmological constant, one can
change the sign of the gravitational constant, but if $\Lambda <0$, this will give a negative mass to the BTZ
black hole, so the existence of a stable ground state is in doubt in this model \cite{5}.  TMG has a bulk-boundary
unitarity conflict. In another term either the bulk or the boundary theory is non-unitary, so there is a clash between the positivity of the two Brown-Henneaux boundary $c$ charges
 and the bulk energies \cite{11}. Recently an interesting three dimensional massive gravity introduced by Bergshoeff, et. al \cite{17} which dubbed Minimal Massive Gravity (MMG), which has the
same minimal local structure as TMG. The MMG model has the same gravitational
degree of freedom as the TMG has and the linearization of the metric field equations for MMG yield a single propagating
massive spin-2 field. It seems that the single massive degree of freedom of MMG is
unitary in the bulk and gives rise to a unitary CFT on the boundary.\\
 The authors of \cite{1'} by formulating TMG at a special point in parameter space, where the curvature radius of $AdS_3$ equals the inverse of the graviton mass $\mu$, have introduced the Chiral Gravity(CG). Recently the authors of \cite{1} have proposed a generalization of chiral gravity. They have considered a Chern-Simons action for the spin connection in the presence of a scalar field and a constraint that enforce the spin connection remain torsion-less. So, the model includes TMG and CG as particular cases. Here we would like extend the Lagrangian of this model such that it describes minimal massive gravity theory non-minimally coupled to a scalar field. Then we try to find a general formula for quasi-local conserved charge of the model in first order formalism. We use the obtained formula to obtain energy, angular momentum and entropy of the BTZ black hole solution in the context of this theory.\\
 There are several approach to obtain mass and angular momentum of black holes solutions  of different gravity theories \cite{2'}-\cite{12'}. The authors of \cite{6} have obtained the quasi-local conserved charges for  black holes in any diffeomorphically invariant theory of gravity. By considering an appropriate variation of the metric, they have established a one-to-one correspondence between the ADT approach and the linear Noether expressions. They have extended this work to a theory of gravity containing a gravitational Chern-Simons term in \cite{2''}, and have computed the off-shell potential and quasi-local conserved charges of some black holes in TMG.\\
  In the metric formalism of gravity for the covariant theories defined by a Lagrangian $n$-form $L$, Wald showed that the entropy of black holes is the Noether charge associated with the horizon-generating Killing vector field, evaluated at the bifurcation surface \cite{131}. Presence of the purely gravitational Chern-Simons terms and mixed gauge gravitational ones gives rise to a non-covariant theory of gravity in the metric formalism.
  Tachikawa extended the Wald approach to include non-covariant theories\cite{141}. By this extension one can obtain the black hole entropy as a Noether charge in the context of non-covariant theories as well. This extension was
then worked out by authors of \cite{142}.  The authors of \cite{143} have proposed a new formulation of a differential Noether charge for theories
in the presence of Chern-Simons terms. They have also presented a manifestly
covariant derivation of the Tachikawa formula for Chern-Simons contribution to entropy.\\
The remainder of this paper is organized as follows. In section 2, at first we shortly introduce the model of \cite{1}, then by adding a convenient term to the Lagrangian of \cite{1} we generalize this model  such that it describes minimal massive gravity theory non-minimally coupled to a scalar field. In contrast to the the Lagrangian of \cite{1}, our model is not torsion free. In section 3, we obtain the equations of motion. We show that the new filed $h$, in contrast to the usual minimal massive gravity is not a symmetric tensor. Also we show that the BTZ black hole spacetime solves the equations of motion. In section 4, we will find an expression to conserved charges of considered model associated with asymptotic Killing vector field $\xi$ based on quasi-local formalism for conserved charges. In section 5 we consider a stationary black hole solution of the minimal massive gravity coupled to scalar field, then we find a general formula for entropy of such black hole solutions. In section 6 we apply our obtained formula for conserved charges and entropy on the BTZ black hole solution of minimal massive gravity coupled to scalar field model, and obtain energy, angular momentum and entropy of these black holes. In section 7, we calculate the central extension term and, through it, we read off the central charges and the eigenvalues of the Virasoro algebra generators for the BTZ black hole solution. Also we obtain again the energy and angular momentum of this black hole using the eigenvalues of the Virasoro algebra generators. Then we find the entropy of BTZ black hole by using the Cardy formula. In the final section, we summarize our results.
\section{Minimal massive gravity non-minimally coupled to a scalar filed}
Recently, it has been shown that one can achieves from Lorentz-Chern-Simons action to topologically massive gravity (TMG) non-minimally coupled to a scalar field \cite{1}. We know that the Lorentz-Chern-Simons Lagrangian 3-form is given by
\begin{equation}\label{1}
  L_{CS}(\omega)= \omega ^{a}_{\hspace{1.5 mm} b} \wedge d \omega ^{b}_{\hspace{1.5 mm} a} + \frac{2}{3} \omega ^{a}_{\hspace{1.5 mm} b} \wedge \omega ^{b}_{\hspace{1.5 mm} c} \wedge \omega ^{c}_{\hspace{1.5 mm} a},
\end{equation}
where $\omega ^{ab} = \omega ^{ab}_{\hspace{2.5 mm} \mu} dx^{\mu}$ are the components of spin-connection 1-form. \footnote{Here, we use Latin and Greek Letters to characterize Lorentz and coordinate indices, respectively.} We can decompose the spin-connection in two independent parts
\begin{equation}\label{2}
  \omega ^{ab} = \tilde{\omega} ^{ab} + K^{ab},
\end{equation}
where $\tilde{\omega} ^{ab}$ is the torsion-free part which is known as Riemannian spin-connection and $K^{ab}$ is contorsion 1-form. The field equations for the Lorentz-Chern-Simons Lagrangian are
\begin{equation}\label{3}
  R^{ab}(\omega)= d \omega ^{ab} + \omega ^{a}_{\hspace{1.5 mm} c} \wedge \omega ^{cb} =0,
\end{equation}
where $R^{ab}(\omega)$ is curvature 2-form. Using following the Bianchi identities
\begin{equation}\label{4}
  D(\omega) R^{ab}(\omega) = 0 , \hspace{1.5 cm} D(\omega) T^{a}(\omega) = R ^{a}_{\hspace{1.5 mm} b}(\omega) \wedge e^{b},
\end{equation}
we find that $D(\omega) T^{a}(\omega) =0 $, and this equation has the following solution in three dimensions
\begin{equation}\label{5}
  T^{a}(\omega)= \varphi _{0} \varepsilon ^{a} _{\hspace{1.5 mm} bc} e^{b} \wedge e^{c},
\end{equation}
where $\varphi _{0}$ is a constant. Notice that $D(\omega)$ denotes exterior covariant derivative with respect to $\omega$ and $T^{a}(\omega)$ is torsion 2-form which is defined as
\begin{equation}\label{6}
  T^{a}(\omega) = D(\omega) e^{a} = d e^{a} + \omega ^{a} _{\hspace{1.5 mm} b} \wedge e^{b},
\end{equation}
also, $e^{a} = e^{a} _{\hspace{1.5 mm} \mu} dx^{\mu}$ are vector valued 1-forms such that $e^{a} _{\hspace{1.5 mm} \mu}$ denote dreibein and we shall assume that the dreibein is invertible. Since $ T^{a}(\omega) = K^{a} _{\hspace{1.5 mm}b} \wedge e^{b}$, then we find that
\begin{equation}\label{7}
  K^{ab} = - \varphi _{0} \varepsilon ^{ab} _{\hspace{2.5 mm} c} e^{c}.
\end{equation}
In the paper \cite{1}, the authors promote $\varphi _{0}$ to be a local dynamical field $\varphi = \varphi(x)$. By substituting Eq.\eqref{2} with $K^{ab} = - \varphi \varepsilon ^{ab} _{\hspace{2.5 mm} c} e^{c}$ into the Lagrangian \eqref{1} we have
\begin{equation}\label{8}
\begin{split}
   L_{CS}(\omega) = & \frac{1}{2} L_{CS}(\tilde{\omega}) + \varphi \varepsilon _{abc} e^{a} \wedge R^{bc}(\tilde{\omega}) + \frac{1}{3} \varphi ^{3} \varepsilon _{abc} e^{a} \wedge e^{b} \wedge e^{c} \\
     & + \varphi ^{2} e_{a} \wedge T^{a} (\tilde{\omega}) + \frac{1}{2} d \left( \varphi \varepsilon _{abc} \tilde{\omega} ^{ab} \wedge e^{c} \right).
\end{split}
\end{equation}
Eventually, the authors of the paper \cite{1} presented the following Lagrangian
\begin{equation}\label{9}
\begin{split}
   L_{[\lambda , m]} = & \varphi \varepsilon _{abc} e^{a} \wedge R^{bc}(\tilde{\omega}) + \frac{1}{3!} \lambda \varphi ^{3} \varepsilon _{abc} e^{a} \wedge e^{b} \wedge e^{c} + \frac{1}{2m} L_{CS}(\tilde{\omega}) \\
     & + \varphi ^{2} e_{a} \wedge T^{a} (\tilde{\omega}) + \frac{1}{2} d \left( \varphi \varepsilon _{abc} \tilde{\omega} ^{ab} \wedge e^{c} \right) + \frac{1}{2m} \zeta _{a} \wedge T^{a} (\tilde{\omega}),
\end{split}
\end{equation}
where $\lambda$ and $m$ are two parameters and they are introduced to adapt cosmological constant and mass parameter of TMG term, respectively. The last term in the Lagrangian \eqref{9} makes this theory to be torsion free. The above Lagrangian describes a Topologically massive gravity theory non-minimally coupled to a scalar field.\\
In thee dimensions, it is convenient to define dualized spin connection 1-form and dualized curvature 2-form as follows:
\begin{equation}\label{10}
  \omega ^{a}= \frac{1}{2} \varepsilon ^{a}_{\hspace{1.5 mm} bc} \omega ^{bc}, \hspace{1.5 cm} R ^{a}= \frac{1}{2} \varepsilon ^{a}_{\hspace{1.5 mm} bc} R ^{bc},
\end{equation}
respectively. By using a 3D-vector algebra notation for Lorentz vectors (see for instance \cite{2}), dualized curvature and torsion 2-forms can be written as
\begin{equation}\label{11}
  R=d \omega + \frac{1}{2} \omega \times \omega , \hspace{1.5 cm} T = de+ \omega \times e,
\end{equation}
in terms of dualized spin-connection, respectively.\\
Now, we want to generalize the Lagrangian \eqref{9} such that it describes Minimal massive gravity theory non-minimally coupled to a scalar field. First of all, we consider the following redefinitions in the Lagrangian \eqref{9}
\begin{equation}\label{12}
\begin{split}
     & \lambda \rightarrow 2 \lambda, \hspace{1 cm} m \rightarrow \mu , \hspace{1 cm}  L_{[\lambda , m]} \rightarrow  2L, \\
     & \hspace{1 cm} \frac{1}{4m} \zeta \rightarrow h , \hspace{1 cm} \tilde{\omega} \rightarrow \omega,
\end{split}
\end{equation}
and then, we add the following term to the obtained Lagrangian
\begin{equation}\label{13}
  \frac{1}{2} \alpha e \cdot h \times h ,
\end{equation}
where $\alpha$ is just a dimensionless parameter. Thus, we obtain the following Lagrangian
\begin{equation}\label{14}
\begin{split}
   L = & \varphi e \cdot R(\omega) + \frac{1}{3!} \lambda \varphi ^{3} e \cdot e \times e + \frac{1}{2 \mu} ( \omega \cdot d \omega + \frac{1}{3} \omega \cdot \omega \times \omega ) \\
     & + \frac{1}{2} \varphi ^{2} e \cdot T (\omega) + \frac{1}{2} d \left( \varphi \omega \cdot e \right) + h \cdot T (\omega) + \frac{1}{2} \alpha e \cdot h \times h.
\end{split}
\end{equation}
It is easy to see that the theory described by the above Lagrangian is not a torsion free one.
\section{Equations of motion}
To find equations of motion, we must take variation of the Lagrangian \eqref{14}
\begin{equation}\label{15}
     \delta L = \delta \varphi E _{\varphi} + \delta e \cdot E _{e} + \delta \omega \cdot E _{\omega} + \delta h \cdot E _{h} + d \Theta (\Phi ,  \delta \Phi)
\end{equation}
where $\Phi$ is collection of all the fields, i.e. $\Phi = \{ \varphi , e, \omega , h \}$. In the above equation, we have the following definitions
\begin{equation}\label{16}
  E_{\varphi} = e \cdot R (\omega) + \frac{\lambda}{2} \varphi ^{2} e \cdot e \times e + \varphi e \cdot T(\omega),
\end{equation}
\begin{equation}\label{17}
  E_{e} = \varphi R (\omega) + \frac{\lambda}{2} \varphi ^{3} e \times e + \frac{1}{2} \varphi ^{2} T(\omega) + \frac{1}{2} \alpha h \times h + \frac{1}{2} D (\omega) \left( \varphi ^{2} e \right) + D (\omega) h ,
\end{equation}
\begin{equation}\label{18}
  E_{\omega} = \frac{1}{\mu} R(\omega) + \frac{1}{2} \varphi ^{2} e \times e + e \times h + D (\omega) \left( \varphi e \right),
\end{equation}
\begin{equation}\label{19}
  E_{h} = T(\omega) + \alpha e \times h ,
\end{equation}
\begin{equation}\label{20}
  \Theta (\Phi , \delta \Phi) = \varphi \delta \omega \cdot e + \frac{1}{2 \mu} \delta \omega \cdot \omega + \frac{1}{2} \varphi ^{2} \delta e \cdot e + \delta e \cdot h + \frac{1}{2} \delta (\varphi \omega \cdot e).
\end{equation}
The equations of motion of the considered theory are
\begin{equation}\label{21}
  E _{\varphi} = E _{e} = E _{\omega} = E _{h} =0,
\end{equation}
and $\Theta (\Phi , \delta \Phi)$ is just surface term. We can write the equation of motion \eqref{19}, namely $E_{h}= 0$, as
\begin{equation}\label{22}
  T(\omega) = d e + (\omega + \alpha h) \times e = 0,
\end{equation}
It is clear that one can define a new dual spin-connection 1-form
\begin{equation}\label{23}
  \Omega = \omega + \alpha h,
\end{equation}
which is the usual torsion free spin-connection $\Omega = \Omega (e)$. By this definition, the equations of motion can be rewritten as
\begin{equation}\label{24}
   e \cdot R (\Omega) + \frac{\lambda}{2} \varphi ^{2} e \cdot e \times e - \alpha \varphi e \cdot e \times h + \frac{1}{2} \alpha ^{2} e \cdot h \times h - \alpha e \cdot D(\Omega) h =0,
\end{equation}
\begin{equation}\label{25}
  \varphi R (\Omega) + \frac{\lambda}{2} \varphi ^{3} e \times e - \alpha \varphi ^{2} e \times h - \frac{1}{2} \alpha ( 1 - \alpha \varphi ) h \times h + ( 1 - \alpha \varphi ) D (\Omega) h + \varphi d \varphi \hspace{0.5 mm} e =0,
\end{equation}
\begin{equation}\label{26}
   R(\Omega) + \frac{1}{2} \mu \varphi ^{2} e \times e + \mu ( 1 - \alpha \varphi ) e \times h - \alpha D (\Omega) h + \frac{1}{2} \alpha ^{2} h \times h + \mu  d \varphi \hspace{0.5 mm} e =0,
\end{equation}
\begin{equation}\label{27}
  T(\Omega) = 0 .
\end{equation}
To obtain the above equations we have used the following equation
\begin{equation}\label{28}
  D (\omega) f = D (\Omega) f - \alpha h \times f,
\end{equation}
where $f$ is an arbitrary Lorentz vector valued 1-form. By combining the equations \eqref{25} and \eqref{26}, we have
\begin{equation}\label{29}
\begin{split}
     & R(\Omega) + \frac{1}{2} \left[ \alpha \lambda \varphi + \mu (1 - \alpha \varphi) \right] \varphi ^{2} e \times e + \left[ \mu (1 - \alpha \varphi) ^{2} - \alpha ^{2} \varphi ^{2} \right] e \times h \\
     & + \left[ \alpha \varphi + \mu (1 - \alpha \varphi) \right] d \varphi \hspace{0.5 mm} e = 0.
\end{split}
\end{equation}
We can solve this equation to find the following expression for $h$,
\begin{equation}\label{30}
\begin{split}
   h^{a} _{\hspace{1.5 mm} \mu} = - \frac{1}{\left[ \mu (1 - \alpha \varphi) ^{2} - \alpha ^{2} \varphi ^{2} \right] } & \{ S ^{a} _{\hspace{1.5 mm} \mu} + \frac{1}{2} \left[ \alpha \lambda \varphi + \mu (1 - \alpha \varphi) \right] \varphi ^{2} e ^{a} _{\hspace{1.5 mm} \mu} \\
     & \left[ \alpha \varphi + \mu (1 - \alpha \varphi) \right] \varepsilon ^{ab}_{\hspace{3 mm} c} e _{b}^{\hspace{1.5 mm} \nu} e^{c} _{\hspace{1.5 mm} \mu} \partial _{\nu} \varphi \}.
\end{split}
\end{equation}
In contrast to ordinary minimal massive gravity \cite{17}, $h_{\mu \nu}$ is not a symmetric tensor, i.e. in the considered model the condition $e \cdot h =0$ no longer holds. In the equation \eqref{30}, $S_{\mu \nu } = \mathcal{R} _{\mu \nu} - \frac{1}{4} g_{\mu \nu} \mathcal{R}$ is 3D Schouten tensor, where $\mathcal{R} _{\mu \nu}$ and $\mathcal{R}$ are respectively Ricci tensor and Ricci scalar.\\
For BTZ black hole spacetime \cite{4''}, we have
\begin{equation}\label{31}
  R(\Omega) = - \frac{1}{2 l^{2}} e \times e , \hspace{1 cm} S^{a} = - \frac{1}{2 l^{2}} e,
\end{equation}
where $l$ is AdS space radii. By assuming that $\varphi$ be a constant, say $\varphi = \varphi _{0}$, the BTZ black hole spacetime solves the equations of motion \eqref{24}-\eqref{27}. So, by taking $\varphi = \varphi _{0}$, for BTZ black hole spacetime the equation \eqref{30} reduces to
\begin{equation}\label{32}
  h^{a}= \beta e ^{a},
\end{equation}
where
\begin{equation}\label{33}
  \beta = \frac{1 - \alpha \lambda l^{2} \varphi _{0} ^{3} - \mu l^{2} (1 - \alpha \varphi _{0}) \varphi _{0} ^{2}}{2l^{2}\left[ \mu (1 - \alpha \varphi _{0}) ^{2} - \alpha ^{2} \varphi _{0} ^{2} \right]}.
\end{equation}
By substituting Eq.\eqref{31} and Eq.\eqref{32} into the equations of motion \eqref{24}-\eqref{27}, we have
\begin{equation}\label{34}
  -\frac{1}{2 l^{2}} + \frac{1}{2} \lambda \varphi _{0} ^{2} - \alpha \beta \varphi _{0} + \frac{1}{2} \alpha ^{2} \beta ^{2} =0,
\end{equation}
\begin{equation}\label{35}
  -\frac{\varphi _{0}}{2 l^{2}} + \frac{1}{2} \lambda \varphi _{0} ^{3} - \alpha \beta \varphi _{0} ^{2} - \frac{1}{2} \alpha \beta ^{2} (1 - \alpha \varphi _{0}) =0,
\end{equation}
\begin{equation}\label{36}
  -\frac{1}{2 l^{2}} + \frac{1}{2} \mu \varphi _{0} ^{2} + \mu \beta (1 - \alpha \varphi _{0}) + \frac{1}{2} \alpha ^{2} \beta ^{2} =0.
\end{equation}
It is obvious that by combining Eq.\eqref{35} and Eq.\eqref{36}, the equation \eqref{33} can be regained. Thus, the BTZ black hole spacetime together with $\varphi = \varphi _{0}$ will be a solution of the considered model when the equation \eqref{34} is satisfied, where $\beta$ is given by Eq.\eqref{33}.
\section{Quasi-local conserved charges}
In this section, we will find an expression to conserved charges of considered model associated with asymptotic Killing vector field $\xi$ based on quasi-local formalism for conserved charges \cite{6,2'',70,80,90}.\\
Under Lorentz gauge transformation $\Lambda \in SO(2,1) $, dreibein transforms as $e ^{a}_{\hspace{1.5 mm} \mu} \rightarrow \Lambda ^{a}_{\hspace{1.5 mm} b} e^{b}_{\hspace{1.5 mm} \mu}$ so that the spacetime metric $g_{\mu \nu}=\eta _{ab} e^{a}_{\hspace{1.5 mm} \mu} e^{b}_{\hspace{1.5 mm} \nu}$ under this transformation remains unchanged. Also, under Lorentz gauge transformation the spin-connection transforms as $\omega \rightarrow \Lambda \omega \Lambda ^{-1}+ \Lambda d \Lambda ^{-1}$ so this is not an invariant quantity under considered transformation. One can define Lorentz-Lie (L-L) derivative of the dreibein 1-form as \cite{101}
\begin{equation}\label{37}
  \mathfrak{L}_{\xi} e^{a} = \pounds_{\xi} e^{a} +\lambda ^{a}_{\hspace{1.5 mm} b} e^{b}.
\end{equation}
where $\pounds_{\xi}$ denotes ordinary Lie derivative along $\xi$ and $\lambda ^{a}_{\hspace{1.5 mm} b}$ generates the Lorentz gauge transformations $SO(2,1)$. In general, $\lambda ^{a}_{\hspace{1.5 mm} b}$ is independent of the dynamical fields of considered model and it is a function of spacetime coordinates and of the diffeomorphism generator $\xi$. The total variation of the dreibein and the spin-connection are defined as \cite{111}
\begin{equation}\label{38}
  \delta _{\xi} e^{a} = \mathfrak{L}_{\xi} e^{a},
\end{equation}
 \begin{equation}\label{39}
   \delta _{\xi} \omega = \mathfrak{L}_{\xi} \omega -d \chi _{\xi},
 \end{equation}
respectively, where $\chi^{a} _{\xi} = \frac{1}{2} \varepsilon ^{a}_{\hspace{1.5 mm} bc} \lambda ^{bc}$. The extra term in \eqref{39}, $-d \chi _{\xi}$, can makes a theory non-covariant, in the meaning of Lorentz covariance (because $e$ and $\omega$ both are invariant under general coordinate transformation).\\
Now, we suppose that the variation of the Lagrangian Eq.\eqref{15} is due to a diffeomorphism which is generated by the vector field $\xi$, so the total variation of Lagrangian \eqref{15} with respect to the diffeomorphism $\xi$ is
\begin{equation}\label{40}
     \delta _{\xi} L = \delta _{\xi} \varphi E _{\varphi} + \delta _{\xi} e \cdot E _{e} + \delta _{\xi} \omega \cdot E _{\omega} + \delta _{\xi} h \cdot E _{h} + d \Theta (\Phi ,  \delta _{\xi} \Phi).
\end{equation}
On the one hand, presence of topological Chern-Simons term in the Lagrangian \eqref{14} makes this model to be Lorentz non-covariant, by virtue of Eq.\eqref{39}. So, the total variation of the Lagrangian \eqref{14} due to diffeomorphism generator $\xi$ can be written as
\begin{equation}\label{41}
  \delta _{\xi} L = \mathfrak{L}_{\xi} L +  d \psi _{\xi}.
\end{equation}
On the other hand, from definition of total variation due to $\xi$, equations \eqref{38} and \eqref{39}, we can write
\begin{equation}\label{42}
  \delta _{\xi} e = D(\omega) i_{\xi} e + i_{\xi} T(\omega) + (\chi _{\xi} -  i_{\xi} \omega) \times e ,
\end{equation}
\begin{equation}\label{43}
  \delta _{\xi} \omega =  i_{\xi} R(\omega) + D(\omega) ( i_{\xi} \omega - \chi _{\xi}) ,
\end{equation}
\begin{equation}\label{44}
  \delta _{\xi} h = D(\omega) i_{\xi} h + i_{\xi} D(\omega) h + (\chi _{\xi} -  i_{\xi} \omega) \times h ,
\end{equation}
\begin{equation}\label{45}
  \delta _{\xi} \varphi = i_{\xi} D(\omega) \varphi ,
\end{equation}
where $i_{\xi}$ denotes the interior product in $\xi$. By substituting equations \eqref{41}-\eqref{45} into Eq.\eqref{40} we have
\begin{equation}\label{46}
\begin{split}
     & d \left[ \Theta (\Phi ,  \delta _{\xi} \Phi) - i_{\xi} L - \psi _{\xi} + i_{\xi} e \cdot E _{e} + (i_{\xi} \omega - \chi _{\xi} ) \cdot E _{\omega} + i_{\xi} h \cdot E _{h} \right] = \\
     & (i_{\xi} \omega - \chi _{\xi} ) \cdot \left[ D(\omega) E_{\omega} + e \times E_{e} + h \times E_{h} \right] + i_{\xi} e \cdot D(\omega) E_{e} + i_{\xi} h \cdot D(\omega) E_{h} \\
     & - i_{\xi} T(\omega) \cdot E_{e} - i_{\xi} R(\omega) \cdot E_{\omega} - i_{\xi} D(\omega) h \cdot E_{h} - E_{\varphi} i_{\xi} D(\omega)\varphi .
\end{split}
\end{equation}
The right hand side of above equation becomes zero by virtue of the Bianchi identities \eqref{4}. Therefore, we find that
\begin{equation}\label{47}
  d J _{\xi} =0,
\end{equation}
where
\begin{equation}\label{48}
  J_{\xi} = \Theta (\Phi ,  \delta _{\xi} \Phi) - i_{\xi} L - \psi _{\xi} + i_{\xi} e \cdot E _{e} + (i_{\xi} \omega - \chi _{\xi} ) \cdot E _{\omega} + i_{\xi} h \cdot E _{h}.
\end{equation}
Thus, the quantity $J_{\xi}$ defined above is conserved off-shell. Because $J_{\xi}$ is closed, then by virtue of the Poincare lemma, it is exact such that we can write $J_{\xi} =  d K _{\xi}$. Since this model is not Lorentz covariant we expect that the total variation of surface term differs from its L-L derivative
\begin{equation}\label{49}
  \delta _{\xi} \Theta (\Phi , \delta \Phi) =  \mathfrak{L}_{\xi} \Theta (\Phi , \delta \Phi) + \Pi _{\xi}.
\end{equation}
Now, we take an arbitrary variation from Eq.\eqref{48} and we find that
\begin{equation}\label{50}
  \mathfrak{J}_{ADT}(\Phi , \delta \Phi ; \xi) = d \left[ \delta K _{\xi} - i_{\xi} \Theta (\Phi , \delta \Phi) \right] + \delta \psi _{\xi} - \Pi _{\xi},
\end{equation}
where $\mathfrak{J}_{ADT}(\Phi , \delta \Phi ; \xi)$ is defined as
\begin{equation}\label{51}
  \begin{split}
     \mathfrak{J}_{ADT}(\Phi , \delta \Phi ; \xi) = & \delta e \cdot i_{\xi} E _{e} + \delta \omega \cdot i_{\xi} E _{\omega} + \delta h \cdot i_{\xi} E _{h} - \delta \varphi i_{\xi} E _{\varphi} \\
       & + i_{\xi}e \cdot \delta E _{e} + (i_{\xi} \omega - \chi _{\xi}) \cdot \delta E_{\omega} + i_{\xi} h \cdot \delta E _{h} \\
       & + \delta \Theta (\Phi , \delta _{\xi} \Phi) - \delta _{\xi} \Theta (\Phi , \delta \Phi),
  \end{split}
\end{equation}
and we will refer to that as "generalized off-shell ADT current" \cite{70,80}. If $\xi$ be a Killing vector field everywhere then the generalized off-shell ADT current is reduced to the ordinary one, because we have the following configuration space result given in \cite{121}
\begin{equation}\label{52}
  \delta \Theta (\Phi , \delta _{\xi} \Phi) - \delta _{\xi} \Theta (\Phi , \delta \Phi)=0,
\end{equation}
this equality holds when $\xi$ is a Killing vector field. Also, if the equations of motion and the linearized equations of motion both satisfied,
then the off-shell ADT current is reduced to the symplectic current, and we know that the symplectic current give us conserved charges associated to asymptotically Killing vector fields \cite{131}. Thus, this generalization makes sense for ordinary ADT current. In this way, generalized off-shell ADT current will be conserved for any asymptotically Killing vector field as well as a Killing vector filed which is admitted by spacetime everywhere. It seems that we can write \cite{2'',141}
\begin{equation}\label{53}
  \delta \psi _{\xi} - \Pi _{\xi} = d Z _{\xi},
\end{equation}
so the equation \eqref{50} can be rewritten as
\begin{equation}\label{54}
  \mathfrak{J}_{ADT}(\Phi , \delta \Phi ; \xi) = d \mathfrak{Q}_{ADT}(\Phi , \delta \Phi ; \xi),
\end{equation}
where $\mathfrak{Q}_{ADT}(\Phi , \delta \Phi ; \xi)$ is generalized off-shell ADT conserved charge associated to asymptotically Killing vector field $\xi$ which is given as
\begin{equation}\label{55}
  \mathfrak{Q}_{ADT}(\Phi , \delta \Phi ; \xi) = \delta K _{\xi} - i_{\xi} \Theta (\Phi , \delta \Phi) + Z _{\xi}.
\end{equation}
The quasi-local conserved charge associated to the Killing vector field can be define as $\xi$ as \cite{6,2''}
\begin{equation}\label{56}
  Q (\xi)= \frac{1}{8 \pi G} \int_{0}^{1} ds \int_{\Sigma} \mathfrak{Q} _{ADT}(\Phi | s ) ,
\end{equation}
where $G$ denotes Newtonian gravitational constant and $\Sigma $ is a space-like co-dimension two surface. Also, integration over $s$ is just integration over an one-parameter path in the solution space and $s=0$ and $s=1$ are correspond to the background solution and the interested solution, respectively.\\
It is straightforward to calculate $\psi _{\xi}$ in Eq.\eqref{41} using the fact that exterior derivative and L-L derivative do not commute
\begin{equation}\label{57}
  \left[ d , \mathfrak{L}_{\xi} \right] e = d \chi _{\xi} \times e.
\end{equation}
Thus, we will have
\begin{equation}\label{58}
  \psi _{\xi} = d \chi _{\xi} \cdot \left[ - \frac{1}{2} \varphi e + \frac{1}{2 \mu} \omega \right].
\end{equation}
In a similar way, we can obtain $\Pi _{\xi}$ in the equation \eqref{49} as
\begin{equation}\label{59}
  \Pi _{\xi} = d \chi _{\xi} \cdot \left[ - \frac{1}{2} \delta \varphi e - \frac{1}{2} \varphi \delta e + \frac{1}{2 \mu} \delta \omega  \right].
\end{equation}
It is easy to see form equations \eqref{53}, \eqref{58} and \eqref{59} that $d Z _{\xi} = 0$ then we can choose $Z _{\xi}$ to be zero. As mentioned earlier we can write $J_{\xi} = d K_{\xi}$ by Poincare lemma, so from Eq.\eqref{48}, we can find $K_{\xi}$ as follows:
\begin{equation}\label{60}
  \begin{split}
     K_{\xi} = & \varphi (i_{\xi} \omega - \chi _{\xi}) \cdot e + \frac{1}{2 \mu} i_{\xi} \omega \cdot \omega - \frac{1}{\mu} \chi _{\xi} \cdot \omega + \frac{1}{2} \varphi ^{2} i_{\xi} e \cdot e \\
       & + i_{\xi} e \cdot h + \frac{1}{2} \varphi i_{\xi} \omega \cdot e - \frac{1}{2} \varphi i_{\xi} e \cdot \omega .
  \end{split}
\end{equation}
Considering the above results, namely equations \eqref{58}-\eqref{60}, and by taking into account Eq.\eqref{23}, one can calculate the generalized ADT conserved charge \eqref{55} as
\begin{equation}\label{61}
  \begin{split}
     \mathfrak{Q}_{ADT}(\Phi , \delta \Phi ; \xi) = & \left[ (i_{\xi} \Omega - \chi _{\xi}) \cdot e - \alpha i_{\xi} h \cdot e + \varphi i_{\xi} e \cdot e \right] \delta \varphi \\
       & + \left[ \varphi (i_{\xi} \Omega - \chi _{\xi}) + \varphi ^{2} i_{\xi} e + ( 1 - \alpha \varphi ) i_{\xi} h \right] \cdot \delta e \\
       & + \left[ \varphi i_{\xi} e + \frac{1}{\mu} (i_{\xi} \Omega - \chi _{\xi}) - \frac{\alpha}{\mu} i_{\xi} h \right] \cdot \delta \Omega \\
       & + \left[ ( 1 - \alpha \varphi ) i_{\xi} e - \frac{\alpha}{\mu} (i_{\xi} \Omega - \chi _{\xi}) + \frac{\alpha ^{2}}{\mu} i_{\xi} h \right] \cdot \delta h .
  \end{split}
\end{equation}
By demanding that $\delta _{\xi} e^{a}= 0$ explicitly when $\xi$ is a Killing vector field, we find the following expression for $\chi_{\xi}$ \cite{101,111}
\begin{equation}\label{62}
  \chi _{\xi} ^{a} = i_{\xi} \omega ^{a} + \frac{1}{2} \varepsilon ^{a}_{\hspace{1.5 mm} bc} e^{\nu b} (i_{\xi} T^{c})_{\nu} + \frac{1}{2} \varepsilon ^{a}_{\hspace{1.5 mm} bc} e^{b}_{\hspace{1.5 mm} \mu} e^{c}_{\hspace{1.5 mm} \nu} \nabla ^{\mu} \xi ^{\nu} .
\end{equation}
Now, we simplify this expression for the considered model which is not torsion free. It is clear that covariant derivative of a vector field $\xi$ is given as
\begin{equation}\label{63}
  \nabla _{\mu} \xi ^{\nu} = \partial _{\mu} \xi ^{\nu} + \Gamma ^{\nu} _{\hspace{1.5 mm} \mu \sigma} \xi ^{\sigma},
\end{equation}
where $\Gamma ^{\alpha} _{\hspace{1.5 mm} \mu \nu}$ is the connection compatible with metric $g_{\mu \nu}$. On the one hand, we can decompose this connection into two parts
\begin{equation}\label{64}
  \Gamma ^{\alpha} _{\hspace{1.5 mm} \mu \nu} = \left\{ ^{\hspace{1.5 mm}\alpha} _{\mu \hspace{2 mm} \nu} \right\} + K ^{\alpha} _{\hspace{1.5 mm} \mu \nu}
\end{equation}
where $\left\{ ^{\hspace{2 mm}\alpha} _{\mu \hspace{2 mm} \nu} \right\}$ is Levi-Civita connection and $K ^{\alpha} _{\hspace{1.5 mm} \mu \nu}$ is contorsion tensor which is defined as
\begin{equation}\label{65}
  K ^{\alpha} _{\hspace{1.5 mm} \mu \nu} = T ^{\alpha} _{\hspace{1.5 mm} \mu \nu} + T ^{\hspace{1.5 mm}\alpha} _{\mu \hspace{2 mm} \nu} + T ^{\hspace{1.5 mm}\alpha} _{\nu \hspace{2 mm} \mu}.
\end{equation}
in terms of Cartan torsion tensor $T^{\alpha} _{\hspace{1.5 mm} \mu \nu} = \frac{1}{2} \left( \Gamma ^{\alpha} _{\hspace{1.5 mm} \mu \nu} - \Gamma ^{\alpha} _{\hspace{1.5 mm} \nu \mu} \right)$. On the other hand, because $h^{a} = \frac{1}{2} \varepsilon ^{abc} h_{bc}$, from the equation of motion \eqref{19} we have
\begin{equation}\label{66}
  T ^{\alpha} _{\hspace{1.5 mm} \mu \nu} = \frac{\alpha}{2} \left( h ^{\alpha} _{\hspace{1.5 mm} \mu \nu} - h ^{\alpha} _{\hspace{1.5 mm} \nu \mu} \right).
\end{equation}
Thus, we can calculate contorsion tensor in terms of $h ^{\alpha} _{\hspace{1.5 mm} \mu \nu}$ and then by substituting obtained result into Eq.\eqref{64} we find that
\begin{equation}\label{67}
  K ^{\alpha} _{\hspace{1.5 mm} \mu \nu} = 2 T ^{\alpha} _{\hspace{1.5 mm} \mu \nu} - \alpha h ^{\alpha} _{\hspace{1.5 mm} \mu \nu}.
\end{equation}
Now, by substituting Eq.\eqref{67} into Eq.\eqref{62} we will find the following expression for $\chi _{\xi}$
\begin{equation}\label{68}
  \chi _{\xi} ^{a} = i_{\xi} \Omega ^{a} + \frac{1}{2} \varepsilon ^{a}_{\hspace{1.5 mm} bc} e^{b}_{\hspace{1.5 mm} \mu} e^{c}_{\hspace{1.5 mm} \nu} \tilde{\nabla} ^{\mu} \xi ^{\nu} ,
\end{equation}
where $\tilde{\nabla}$ denotes covariant derivative with respect to Levi-Civita connection. Hence, to calculate conserved charges of considered solutions by using Eq.\eqref{56}, we can employ the above expression for $\chi _{\xi}$.
\section{General formula for entropy of black holes in minimal massive gravity coupled to scalar field }
Let us consider a stationary black hole solution of the minimal massive gravity coupled to scalar field. We know that entropy of a black hole is conserved charge associated to the Killing horizon generating Killing field $\zeta$ \cite{131}. We take the codimension two surface $\Sigma$ to be the bifurcate surface $\mathcal{B}$. Assuming that $\zeta$ is the Killing vector field which generates the Killing horizon, so we must set $\zeta=0$ on $\mathcal{B}$. Thus, the equation \eqref{61} reduces to
\begin{equation}\label{69}
     \mathfrak{Q}_{ADT}(\Phi , \delta \Phi ; \zeta) = - \delta \left[ \chi _{\zeta} \cdot \left( \varphi e + \frac{1}{\mu} \Omega - \frac{\alpha}{\mu} h \right) \right]
\end{equation}
on the bifurcate surface. Now take $s = 0$ and $s = 1$ correspond to the considered black hole spacetime and the perturbed one, respectively. Therefore, by integrating from \eqref{69} over one-parameter path in the solution space we have
\begin{equation}\label{70}
      \int_{0}^{1} ds \mathfrak{Q}_{ADT}(\Phi , \delta \Phi ; \zeta) = - \chi _{\zeta} \cdot \left[ \varphi e + \frac{1}{\mu} \Omega - \frac{\alpha}{\mu} h \right].
\end{equation}
The bifurcate surface of a stationary black hole is a circle. In the paper \cite{111}, the authors have shown that on the bifurcate surface we can write
\begin{equation}\label{71}
  \chi ^{a} _{\zeta} = \kappa N^{a},
\end{equation}
 where $\kappa$ is the surface gravity and $N^{a}$ is a vector which is given as
\begin{equation}\label{72}
  N^{\mu} = \left( 0,0, \frac{1}{\sqrt{g_{\phi \phi}}} \right).
\end{equation}
Hence, by substituting Eq.\eqref{70} into Eq.\eqref{56} we will have
\begin{equation}\label{73}
  Q(\zeta) = - \frac{\kappa}{8 \pi G} \int_{0}^{2 \phi} \frac{d \phi}{\sqrt{g_{\phi \phi}}} \left[ \varphi g _{\phi \phi} + \frac{1}{\mu} \Omega _{\phi \phi} - \frac{\alpha}{\mu} h _{\phi \phi} \right].
\end{equation}
which should be calculated on horizon. Now, we can define entropy of a stationary black hole as \cite{131}
\begin{equation}\label{74}
  S = - \frac{2 \pi}{\kappa} Q(\zeta),
\end{equation}
therefore
\begin{equation}\label{75}
  S= \frac{1}{4 G} \int_{0}^{2 \phi} \frac{d \phi}{\sqrt{g_{\phi \phi}}} \left[ \varphi g _{\phi \phi} + \frac{1}{\mu} \Omega _{\phi \phi} - \frac{\alpha}{\mu} h _{\phi \phi} \right],
\end{equation}
which should be calculated on horizon. In the above formula, $h_{\phi \phi}$ is given by
\begin{equation}\label{76}
   h _{\phi \phi} = - \frac{1}{\left[ \mu (1 - \alpha \varphi) ^{2} - \alpha ^{2} \varphi ^{2} \right] } \left\{ S _{\phi \phi} + \frac{1}{2} \left[ \alpha \lambda \varphi + \mu (1 - \alpha \varphi) \right] \varphi ^{2} g _{\phi \phi} \right\}.
\end{equation}
The formula \eqref{75} will be similar to the ordinary minimal massive gravity one when we take $\varphi = \varphi _{0}$.
\section{Application for BTZ black hole solution with $\varphi = \varphi _{0}$}
In this section, we calculate conserved charges and entropy of the BTZ black hole solution with $\varphi = \varphi _{0}$ in the context of the considered model. The following dreibein describes BTZ black hole spacetime
\begin{equation}\label{77}
  \begin{split}
       & e^{0}=\left( \frac{(r^{2}-r_{+}^{2})(r^{2}-r_{-}^{2})}{l^{2}r^{2}} \right)^{\frac{1}{2}}dt \\
       & e^{1}=\left( \frac{l^{2}r^{2}}{(r^{2}-r_{+}^{2})(r^{2}-r_{-}^{2})} \right)^{\frac{1}{2}}dr \\
       & e^{2}=r \left( d \phi -\frac{r_{+}r_{-}}{lr^{2}} dt \right),
  \end{split}
\end{equation}
where $ r_{+} $ and $ r_{-} $ are outer and inner horizon radii of BTZ black hole, respectively. We take the integration surface $\Sigma$ to be a circle with a radius of infinity. Therefore, we can consider the AdS$_{3}$ spacetime to be background corresponds to $s=0$
\begin{equation}\label{78}
   \bar{e}^{0}= \frac{r}{l}dt, \hspace{1 cm} \bar{e}^{1}= \frac{l}{r}dr, \hspace{1 cm} \bar{e}^{2}= r d \phi.
\end{equation}
The bar on a quantity means that the quantity is calculated on background. Thus, the equation \eqref{61} for the considered solution reduces to
\begin{equation}\label{79}
  \begin{split}
     \mathfrak{Q}_{ADT}(\Phi , \delta \Phi ; \xi) = & \left[ \left( \varphi _{0} - \frac{\alpha \beta}{\mu} \right) (i_{\xi} \bar{\Omega} - \bar{\chi} _{\xi}) + \frac{1}{\mu l^{2}} i_{\xi} \bar{e} \right] \cdot \delta e \\
       & + \left[ \left( \varphi _{0} - \frac{\alpha \beta}{\mu} \right) i_{\xi} \bar{e} + \frac{1}{\mu} (i_{\xi} \bar{\Omega} - \bar{\chi} _{\xi}) \right] \cdot \delta \Omega
  \end{split}
\end{equation}
where we have used Eq.\eqref{32} and Eq.\eqref{36}. Quantities are calculated on the background obviously do not depend on parameters of the solution space, so by taking an integration from Eq.\eqref{79} over one-parameter path in the solution space, we have
\begin{equation}\label{80}
  \begin{split}
     \int_{0}^{1} ds \mathfrak{Q}_{ADT}(\Phi , \delta \Phi ; \xi) = & \left[ \left( \varphi _{0} - \frac{\alpha \beta}{\mu} \right) (i_{\xi} \bar{\Omega} - \bar{\chi} _{\xi}) + \frac{1}{\mu l^{2}} i_{\xi} \bar{e} \right] \cdot \Delta e \\
       & + \left[ \left( \varphi _{0} - \frac{\alpha \beta}{\mu} \right) i_{\xi} \bar{e} + \frac{1}{\mu} (i_{\xi} \bar{\Omega} - \bar{\chi} _{\xi}) \right] \cdot \Delta \Omega,
  \end{split}
\end{equation}
where $\Delta \Phi = \Phi _{(s=1)} - \Phi _{(s=0)}$. By substituting Eq.\eqref{80} into Eq.\eqref{56} we find that
\begin{equation}\label{81}
\begin{split}
   Q(\xi) = \frac{1}{8 \pi G} \lim_{r \rightarrow \infty} \int_{0}^{2 \pi} & \{ \left[ \left( \varphi _{0} - \frac{\alpha \beta}{\mu} \right) (i_{\xi} \bar{\Omega} - \bar{\chi} _{\xi}) + \frac{1}{\mu l^{2}} i_{\xi} \bar{e} \right] \cdot \Delta e _{\phi} \\
     & + \left[ \left( \varphi _{0} - \frac{\alpha \beta}{\mu} \right) i_{\xi} \bar{e} + \frac{1}{\mu} (i_{\xi} \bar{\Omega} - \bar{\chi} _{\xi}) \right] \cdot \Delta \Omega _{\phi} \} d \phi.
\end{split}
\end{equation}
For BTZ black hole spacetime at spatial infinity we have
\begin{equation}\label{82}
    \Delta e^{a} _{\hspace{1.5 mm} \phi} = 0, \hspace{0.5 cm} \Delta \Omega ^{\hat{t}} _{\hspace{1.5 mm} \phi} = - \frac{r_{+}^{2}+r_{-}^{2}}{2 l r} , \hspace{0.5 cm} \Delta \Omega ^{\hat{r}} _{\hspace{1.5 mm} \phi} = 0, \hspace {0.5 cm} \Delta \Omega ^{\hat{\phi}} _{\hspace{1.5 mm} \phi} = - \frac{r_{+} r_{-}}{ l r},
\end{equation}
which are calculated based on Eq.\eqref{77}.\\
Energy is conserved charge corresponds to the Killing vector $\xi = \partial _{t}$, thus
\begin{equation}\label{83}
  E = \frac{1}{8 G} \left[ \left( \varphi _{0} - \frac{\alpha \beta}{\mu}  \right) \frac{r_{+}^{2}+r_{-}^{2}}{l^{2}} - \frac{2 r_{+} r_{-}}{ \mu l^{3}} \right],
\end{equation}
and angular momentum is conserved charge corresponds to the Killing vector $\xi = - \partial _{\phi}$, then
\begin{equation}\label{84}
  j = \frac{1}{8 G} \left[ \left( \varphi _{0} - \frac{\alpha \beta}{\mu}  \right) \frac{2 r_{+} r_{-}}{l} - \frac{r_{+}^{2}+r_{-}^{2}}{\mu l^{2}} \right].
\end{equation}
Since on the horizon of BTZ black hole we have
\begin{equation}\label{85}
  g_{\phi \phi} = r_{+}^{2}, \hspace{1 cm} \Omega _{\phi \phi} = - \frac{r_{+} r_{-} }{l} , \hspace{1 cm} h _{\phi \phi} = \beta r_{+}^{2},
\end{equation}
so by substituting Eq.\eqref{85} into Eq.\eqref{75} we find entropy of BTZ black hole solution as
\begin{equation}\label{86}
  S = \frac{\pi}{2 G} \left[ \left( \varphi _{0} - \frac{\alpha \beta}{\mu}  \right) r_{+} - \frac{r_{-}}{\mu} \right].
\end{equation}
It is straightforward to check that these results satisfy the first law of black hole mechanics, that is
\begin{equation}\label{87}
  \delta E = T_{H} \delta S + \Omega _{H} \delta j,
\end{equation}
where $T_{H}= \frac{\kappa}{2 \pi}$ and $\Omega _{H} = \frac{r_{-}}{l r_{+}}$ are Hawking temperature and angular velocity of horizon, respectively.
\section{Virasoro algebra and the central term}
Using the results of section 4 , we can obtain the central extension term for the considered model and subsequently we can read off the central charges. In this section, we take AdS$_{3}$ spacetime with $\varphi = \varphi _{0}$ as background (see Eq.\eqref{78}) and the integration surface $\Sigma$ to be a circle with a radius of infinity. Two copies of the classical centerless Virasoro algebra, which are known as the Witt algebra, is given by
\begin{equation}\label{88}
    [ \xi _{m} ^{\pm} , \xi _{n} ^{\pm} ] = i (n-m) \xi _{m+n} ^{\pm} , \hspace{1 cm} [ \xi _{m} ^{+} , \xi _{n} ^{-} ] = 0 ,
\end{equation}
where $\xi _{m} ^{\pm}$ ( $ m \in \mathbb{Z}$ ) are the vector fields and they have the following form \cite{16}
\begin{equation}\label{89}
   \xi _{n} ^{\pm} = \frac{1}{2} e^{inx^{\pm}} \left[  l \left( 1-\frac{l^{2} n^{2}}{2r^{2}} \right) \partial _{t} -inr \partial _{r} \pm \left( 1+\frac{l^{2} n^{2}}{2r^{2}} \right) \partial _{\phi } \right] ,
\end{equation}
where $x^{\pm}= t/l \pm \phi$. Also, the square brackets in Eq.\eqref{88} denote the Lie bracket. The central extension term $C (\xi _{m} ^{\pm} , \xi _{n} ^{\pm})$ is given by the following equation \cite{15} (see also related works in \cite{151, 152})
\begin{equation}\label{90}
   \{ Q ( \xi _{m} ^{\pm} ) , Q ( \xi _{n} ^{\pm} ) \}  = Q ( [ \xi _{m} ^{\pm} , \xi _{n} ^{\pm} ] ) +  C (\xi _{m} ^{\pm} , \xi _{n} ^{\pm}) .
\end{equation}
Since the conserved charge \eqref{56} is linear in $ \xi $ then
\begin{equation}\label{91}
 Q ( [ \xi _{m} ^{\pm} , \xi _{n} ^{\pm} ] ) = i (n-m) Q( \xi _{m+n} ^{\pm} ) ,
\end{equation}
on the other hand, we know that
\begin{equation}\label{92}
   \{ Q ( \xi _{m} ^{\pm} ) , Q ( \xi _{n} ^{\pm} ) \}  = \delta _{\xi _{n} ^{\pm}} Q( \xi _{m} ^{\pm} ) ,
\end{equation}
thus, the central extension term will be obtained from the following equation
\begin{equation}\label{93}
  C (\xi _{m} ^{\pm} , \xi _{n} ^{\pm})   = \delta _{\xi _{n} ^{\pm}} Q( \xi _{m} ^{\pm} ) - i (n-m) Q( \xi _{m+n} ^{\pm} ) .
\end{equation}
It is clear from Eq.\eqref{56} that we can write at spatial infinity
\begin{equation}\label{94}
  \delta Q (\xi)= \frac{1}{8 \pi G} \int_{\infty} \mathfrak{Q} _{ADT}(\bar{\Phi}, \delta \Phi ; \xi ).
\end{equation}
Therefore, by taking an integration from Eq.\eqref{94} over one-parameter path in the solution space, we have
\begin{equation}\label{95}
  Q (\xi)= \frac{1}{8 \pi G} \int_{\infty} \mathfrak{Q} _{ADT}(\bar{\Phi}, \Delta \Phi ; \xi ),
\end{equation}
also, from Eq.\eqref{94}, we can easily deduce that
\begin{equation}\label{96}
  \delta _{\xi _{n} ^{\pm}} Q( \xi _{m} ^{\pm} ) = \frac{1}{8 \pi G} \int_{\infty} \mathfrak{Q} _{ADT}(\bar{\Phi}, \delta _{\xi _{n} ^{\pm}} \Phi ; \xi _{m} ^{\pm} ).
\end{equation}
Thus, by substituting Eq.\eqref{95} and Eq.\eqref{96} into Eq.\eqref{93}, we find an expression for the central extension term and consequently we can read off the central charges of the considered model. Because we take AdS$_{3}$ spacetime with $\varphi = \varphi _{0}$ then the equation \eqref{96} can be rewritten as
\begin{equation}\label{97}
  \delta _{\xi _{n} ^{\pm}} Q( \xi _{m} ^{\pm} ) = \frac{1}{8 \pi G} \left( \varphi _{0} \pm \frac{1}{\mu l} -\frac{\alpha \beta}{\mu} \right) \lim_{r \rightarrow \infty} \int_{0}^{2 \pi} i_{\xi _{m} ^{\pm}} \bar{e} \cdot \delta _{\xi _{n} ^{\pm}} A^{\pm} _{\phi} d \phi ,
\end{equation}
where $A^{\pm}$ are connections correspond to the two $SO(2, 1)$ gauge group \cite{171}
\begin{equation}\label{98}
  (A^{\pm})^{a} = \Omega ^{a} \pm \frac{1}{l} e ^{a} .
\end{equation}
In the calculation of the equation \eqref{97} we have used the following equation
\begin{equation}\label{99}
  i_{\xi _{n} ^{\pm}} \bar{\Omega} - \bar{\chi} _{\xi _{n} ^{\pm}} = \pm \frac{1}{l} i_{\xi _{n} ^{\pm}} \bar{e} .
\end{equation}
which can be derived from Eq.\eqref{68} and Eq.\eqref{89}. By using Eq.\eqref{23} and Eq.\eqref{28} we can simplified equations \eqref{42}-\eqref{45} as follows:
\begin{equation}\label{100}
  \delta _{\xi} e = D(\Omega) i_{\xi} e + (\chi _{\xi} -  i_{\xi} \Omega) \times e ,
\end{equation}
\begin{equation}\label{101}
  \delta _{\xi} \Omega =  i_{\xi} R(\Omega) + D(\Omega) ( i_{\xi} \Omega - \chi _{\xi}) ,
\end{equation}
\begin{equation}\label{102}
  \delta _{\xi} h = D(\Omega) i_{\xi} h + i_{\xi} D(\Omega) h + (\chi _{\xi} -  i_{\xi} \Omega) \times h ,
\end{equation}
\begin{equation}\label{103}
  \delta _{\xi} \varphi = i_{\xi} D(\Omega) \varphi ,
\end{equation}
therefore, one can show that
\begin{equation}\label{104}
  \begin{split}
       & \delta _{\xi _{n} ^{\pm}} \left( A ^{\pm} \right) ^{0} _{\hspace{1.5 mm} \phi} = - \frac{i l n^{3}}{2r} e^{inx^{\pm}} , \\
       & \delta _{\xi _{n} ^{\pm}} \left( A ^{\pm} \right) ^{1} _{\hspace{1.5 mm} \phi} = 0 , \\
       & \delta _{\xi _{n} ^{\pm}} \left( A ^{\pm} \right) ^{2} _{\hspace{1.5 mm} \phi} = \pm \frac{i l n^{3}}{2r} e^{inx^{\pm}}.
  \end{split}
\end{equation}
By substituting equations \eqref{104} into Eq.\eqref{97}, we find that
\begin{equation}\label{105}
  \delta _{\xi _{n} ^{\pm}} Q( \xi _{m} ^{\pm} ) = \frac{iln^{3}}{8 G} \left( \varphi _{0} \pm \frac{1}{\mu l} -\frac{\alpha \beta}{\mu} \right) \delta _{m+n,0}.
\end{equation}
Suppose that $\varphi= \varphi _{0}$ and $h = \beta e$, as they are sensible for BTZ balck hole solution, then Eq.\eqref{95} for $\xi = \xi _{m} ^{\pm}$ becomes
\begin{equation}\label{106}
   Q( \xi _{m} ^{\pm} ) = \frac{1}{8 \pi G} \left( \varphi _{0} \pm \frac{1}{\mu l} -\frac{\alpha \beta}{\mu} \right) \lim_{r \rightarrow \infty} \int_{0}^{2 \pi} i_{\xi _{m} ^{\pm}} \bar{e} \cdot \Delta A^{\pm} _{\phi} d \phi .
\end{equation}
By substituting equations \eqref{82} into Eq.\eqref{106}, we find that
\begin{equation}\label{107}
   Q( \xi _{m} ^{\pm} ) = \frac{l}{16 G} \left( \varphi _{0} \pm \frac{1}{\mu l} -\frac{\alpha \beta}{\mu} \right) \left( \frac{r_{+} \mp r_{-}}{ l } \right) ^{2} \delta _{m,0} .
\end{equation}
Now, to find central extension term we must substitute Eq.\eqref{105} and Eq.\eqref{107} into Eq.\eqref{93} then we find that
\begin{equation}\label{108}
   C (\xi _{m} ^{\pm} , \xi _{n} ^{\pm}) = \frac{il}{8 G} \left( \varphi _{0} \pm \frac{1}{\mu l} -\frac{\alpha \beta}{\mu} \right) \left[ n^{3} - \left( \frac{r_{+} \mp r_{-}}{ l } \right) ^{2} n \right] \delta _{m+n,0}.
\end{equation}
To obtain the usual $n$ dependence, that is $n (n ^{2} -1)$, in the the above expression it is sufficient one make a shift on $Q( \xi _{m} ^{\pm} )$ by a constant \cite{181}. Thus, by the following substitution
\begin{equation}\label{109}
  Q (\xi _{n} ^{\pm}) \equiv L^{\pm} _{n}, \hspace{1 cm} \{ Q ( \xi _{m} ^{\pm} ) , Q ( \xi _{n} ^{\pm} ) \} \equiv i [L^{\pm} _{m} , L^{\pm} _{n}],
\end{equation}
the equation \eqref{93} becomes
\begin{equation}\label{110}
  [L^{\pm} _{m} , L^{\pm} _{n}] =(n-m) L^{\pm} _{m+n} + \frac{c_{\pm}}{12} n (n ^{2} -1) \delta _{m+n,0} ,
\end{equation}
where
\begin{equation}\label{111}
  c_{\pm}= \frac{3l}{2G} \left( \varphi _{0} \pm \frac{1}{\mu l} -\frac{\alpha \beta}{\mu} \right),
\end{equation}
are central charges and $ L^{\pm}_{n}$ are generators of Virasoro algebra. So the algebra among the conserved charges is isomorphic to two copies of the Virasoro algebra. We can read off the eigenvalues of the Virasoro generators $ L^{\pm}_{n}$ from Eq.\eqref{107} as
\begin{equation}\label{112}
   l^{\pm}_{n} = \frac{l}{16 G} \left( \varphi _{0} \pm \frac{1}{\mu l} -\frac{\alpha \beta}{\mu} \right) \left( \frac{r_{+} \mp r_{-}}{ l } \right) ^{2} \delta _{m,0} .
\end{equation}
The eigenvalues of the Virasoro generators $ L^{\pm}_{n}$ are related to the energy $E$ and the angular momentum $j$ of the BTZ black hole by the following equations respectively
\begin{equation}\label{113}
   E = \frac{1}{l} ( l^{+} _{0} + l^{-} _{0} ) = \frac{1}{8 G} \left[ \left( \varphi _{0} - \frac{\alpha \beta}{\mu}  \right) \frac{r_{+}^{2}+r_{-}^{2}}{l^{2}} - \frac{2 r_{+} r_{-}}{ \mu l^{3}} \right],
\end{equation}
\begin{equation}\label{114}
   j = l^{-} _{0} - l^{+} _{0} =  \frac{1}{8 G} \left[ \left( \varphi _{0} - \frac{\alpha \beta}{\mu}  \right) \frac{2 r_{+} r_{-}}{l} - \frac{r_{+}^{2}+r_{-}^{2}}{\mu l^{2}} \right].
\end{equation}
Also, to calculate the entropy of the considered black hole one can use the Cardy formula \cite{191,201} (see also \cite{16})
\begin{equation}\label{115}
   S = 2 \pi \sqrt{\frac{c_{+} L^{+} _{0} }{6}} + 2 \pi \sqrt{\frac{c_{-} L^{-} _{0} }{6}} = \frac{\pi}{2 G} \left[ \left( \varphi _{0} - \frac{\alpha \beta}{\mu}  \right) r_{+} - \frac{r_{-}}{\mu} \right].
\end{equation}
By comparing above results, equations \eqref{113}-\eqref{115}, with equations \eqref{83},\eqref{84} and \eqref{86} we see that they are exactly matched.
\section{Conclusion}
In this paper, we considered the theory of topologically massive gravity non-minimally coupled to a scalar field which comes from Lorentz-Chern-Simons theory \cite{1}. The Lagrangian of that theory is given by Eq.\eqref{9} and it is a torsion free one. We have extended that theory by adding an extra term \eqref{13} which makes torsion to be non-zero. The Lagrangian of extended theory is given by Eq.\eqref{14} and it could be an extension of minimal massive gravity such that it is non-minimally coupled to a scalar field. In section 3, we obtained equations of motion \eqref{24}-\eqref{27} of extended theory such that they are expressed in terms of usual torsion free spin-connection \eqref{23}. We showed that BTZ spacetime together with $\varphi= \varphi _{0} $ solves the equations of motion \eqref{24}-\eqref{27} when equations \eqref{34}-\eqref{36} are satisfied. In section 4, we defined generalized off-shell ADT current and we deduced that it is conserved for any asymptotically Killing vector field as well as a Killing vector field which is admitted by spacetime everywhere. Then we used Poincare lemma to define generalized off-shell ADT charge \eqref{55} and consequently we defined quasi-local conserved charge \eqref{55} for the considered theory. In section 5, we found general formula \eqref{75} for entropy of stationary black hole solution in the context of considered theory. We used the obtained formulas to calculate energy \eqref{83}, angular momentum \eqref{84} and entropy \eqref{86} of BTZ black hole solution. These quantities satisfy the first law of black hole mechanics. In section 7, we obtained the central extension term \eqref{108} and then we read off the central charges \eqref{111} and the eigenvalues of the Virasoro algebra generators \eqref{112} for the BTZ black hole solution. We calculated energy \eqref{113} and angular momentum \eqref{114} of this black hole using the eigenvalues of the Virasoro algebra generators. Also we calculated the entropy of BTZ black hole by using the Cardy formula \eqref{115}. By comparing equations \eqref{113}-\eqref{115} with equations \eqref{83},\eqref{84} and \eqref{86} we found that although they have been obtained using two different ways, but they are exactly matched as we expected.
\section{Acknowledgments}
M. R. Setare thank Simón. del. Pino and Gaston. Giribet for helpful comments.

\end{document}